\documentclass[a4paper,11pt]{article}
\usepackage{jinstpub} 
\usepackage{lineno}
\usepackage{lmodern}

\title{\boldmath Design and performance of the prototype gaseous beam monitor with GEM and pixel sensors for the CSR external-target experiment}

\author[a,b]{Hulin Wang,}
\author[c,1]{Xianglun Wei,\note{Corresponding author.}}
\author[a,b,2]{Chaosong Gao,\note{Corresponding author.}}
\author[a,b,3]{Jun Liu,\note{Corresponding author.}}
\author[a,b]{Junshuai Liu,}
\author[d]{Zhen Wang,}
\author[a,b]{Ran Chen,}
\author[a,b]{Bihui You,}
\author[c]{Peng Ma,}
\author[c]{Haibo Yang,}
\author[c]{Chengxin Zhao,}
\author[a,b]{Mingmei Xu,}
\author[a,b]{Shusu Shi,}
\author[a,b]{Guangming Huang,}
\author[a,b]{Feng Liu,}
\author[a,b]{Xiangming Sun}

\affiliation[a]{PLAC, Key Laboratory of Quark \& Lepton Physics (MOE), Central China Normal University, Wuhan, 430079, China}
\affiliation[b]{Hubei Provincial Engineering Research Center of Silicon Pixel Chip \& Detection Technology, Wuhan, 430079, China}
\affiliation[c]{Institute of Modern Physics, Chinese Academy of Sciences, Lanzhou, 730000, China}
\affiliation[d]{School of Physics and Electronic Science, Guizhou Normal University, Guiyang, 550001, China}

\emailAdd{weixl@impcas.ac.cn}
\emailAdd{chaosonggao@ccnu.edu.cn}
\emailAdd{junliu@ccnu.edu.cn}

\abstract{
A gaseous beam monitor utilizing gas electron multiplier (GEM) and pixel sensors is being developed for the Cooling Storage Ring (CSR) External-target Experiment (CEE) at Heavy Ion Research Facility in Lanzhou (HIRFL).
The beam monitor is mainly used to track each beam particle, providing an accurate reconstruction of the primary vertex of the collision.
Two generations of the pixel sensors (named Topmetal-CEE) were produced, with the second generation's performance improving over the first one.
The design and performance of the prototype are described in the paper.
Characterization of the prototype with heavy-ion beams and laser beams are presented, showing a spatial resolution better than 50 $\mu$m and a time resolution better than 15 ns.
}

\keywords{Particle tracking detectors (Gaseous detectors), Time projection chambers, CMOS readout of gaseous detectors, Electronic detector readout concepts (gas, liquid), Micropattern gaseous detectors (MSGC, GEM, THGEM, RETHGEM, MHSP, MICROPIC, MICROMEGAS, InGrid, etc)}

\begin{document}
\maketitle
\flushbottom

\section{Introduction}
\label{sec:intro}

Heavy-ion collision experiments worldwide provide controlled tests of quantum chromodynamics~\cite{annurev-nucl-101917-020852}.
The Cooling Storage Ring (CSR)~\cite{XIA200211} External-target Experiment (CEE)~\cite{Lu2016} at Heavy Ion Research Facility in Lanzhou (HIRFL) is a fixed-target experiment currently under development.
Using beams with ion species from H to U and energies from about 0.4 to 1.1 GeV/u provided by the HIFRL-CSR, CEE aims to study the physics of nuclear matter at high baryon density region.

In the CEE spectrometer, 
a dipole magnet provides an uniform magnetic field of 0.5 T.
The tracking of charged particles is provided by a double-volume time projection chamber (TPC)~\cite{Huang2018} and three layers of multi-wire drift chambers (MWDCs)~\cite{mwdc2020}.
The time-of-flight (TOF) system, which is mainly used for the particle identification, is composed of a T0 detector~\cite{t02020}, an inner TOF~\cite{Wang_2022} covering the central rapidity region, and an external TOF~\cite{Wang_2023} for the endcap region.
A zero-degree calorimeter (ZDC)~\cite{9927152}, which measures the collision centrality, is located in the forward region of the spectrometer.

To improve the precision of primary vertex reconstruction, a gaseous beam monitor~\cite{Wang2022} with pixel readout is being developed.
It is to be positioned upstream of the target, measuring the position and time of each beam particle in the offline analysis.
It also monitors the beam status in the online mode.
The main design specifications include a spatial resolution lower than 50 $\mu$m in the directions transverse to the beam, and a time resolution lower than 1 $\mu$s.

\section{Beam Monitor of CEE}
\label{sec:bm}

\begin{figure}[htbp]
    \centering
    \includegraphics[width=0.45\linewidth]{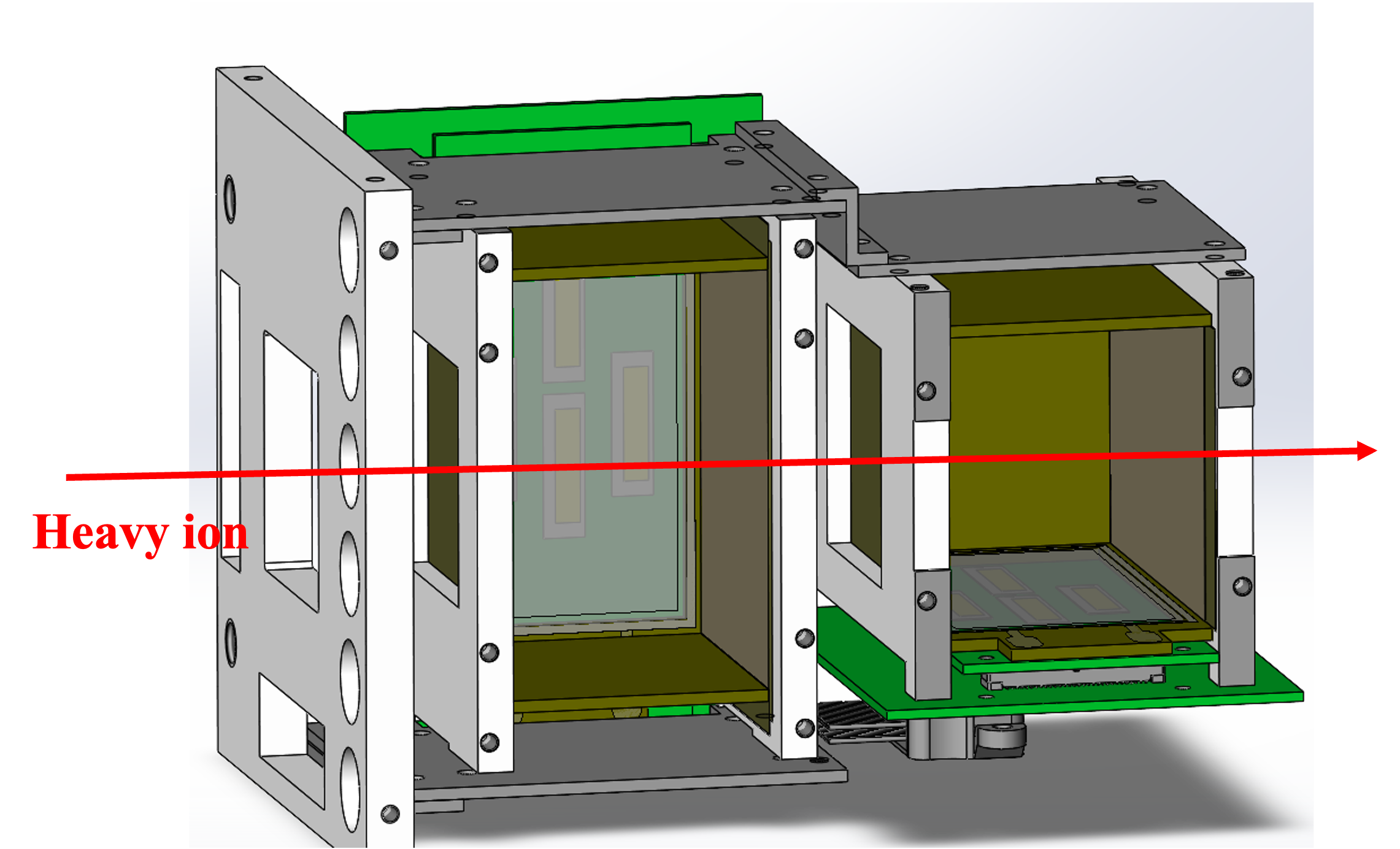}
    \quad
    \includegraphics[width=0.45\linewidth]{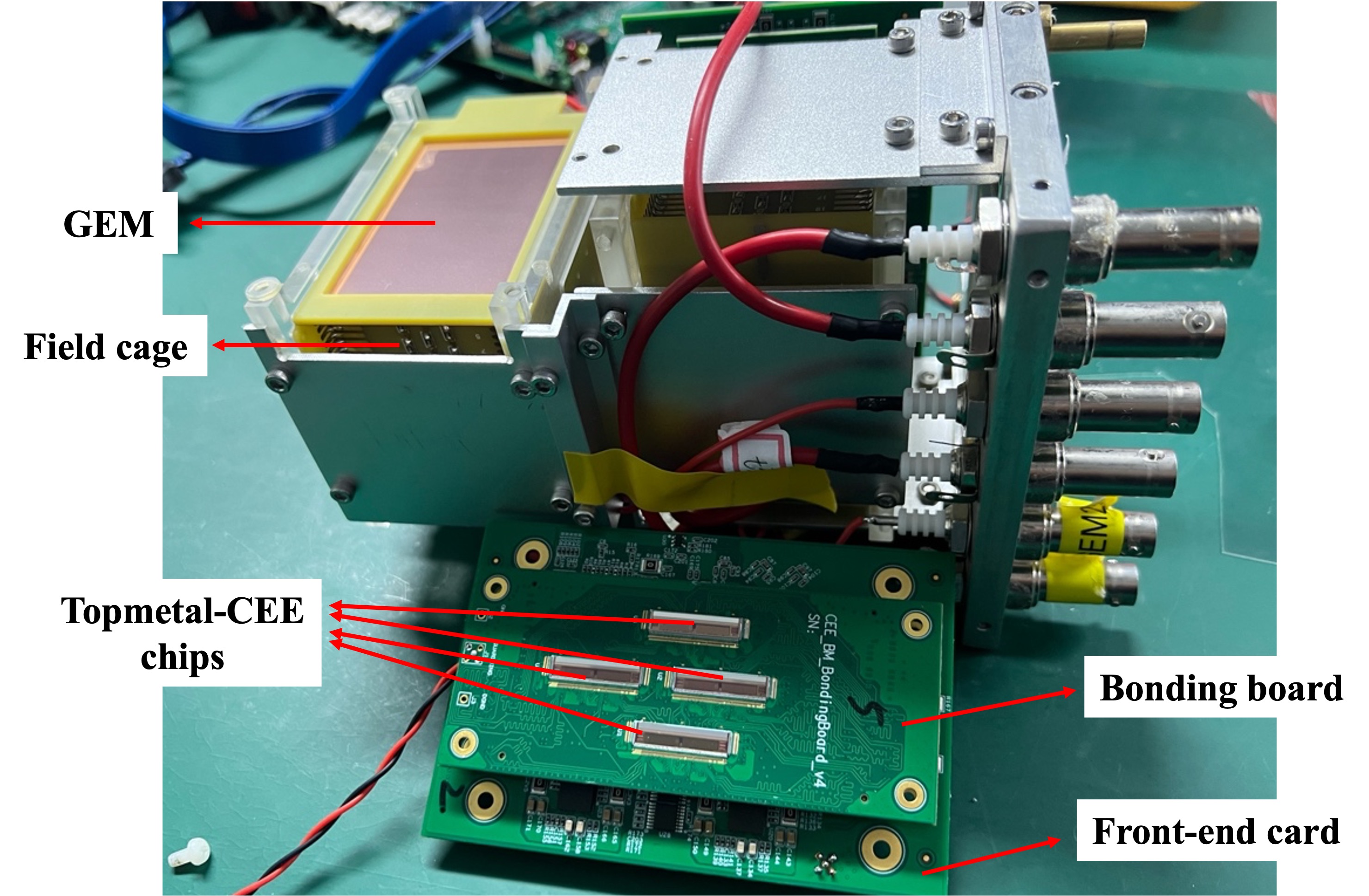}
    \caption{\label{fig:det} The structure (left) and a photograph (right) of the beam monitor. 
             For illustration purpose, the Topmetal-CEE chips and the associated boards are moved from their installed places.}
\end{figure}

The structure of the beam monitor is shown in Figure~\ref{fig:det} (left). 
It has a size of 120 mm $\times$ 120 mm $\times$ 212 mm (along the beam direction).
It contains two micro-TPCs in a gas vessel, with electric field orthogonal to each other.
In this way, the particle positions transverse to the beam direction are measured only using spatial information in the readout planes, 
while the time information is used to combine the measurements in the two micro-TPCs.
Custom-developed Topmetal-CEE chips~\cite{LIU2023167786} are used as direct anode readout.
Gas electron multiplier (GEM) is used for the gas amplification.

\section{Topmetal-CEE charge sensors}
\label{sec:topmetalcee}

\begin{figure}[htbp]
    \centering
    \includegraphics[width=0.45\linewidth]{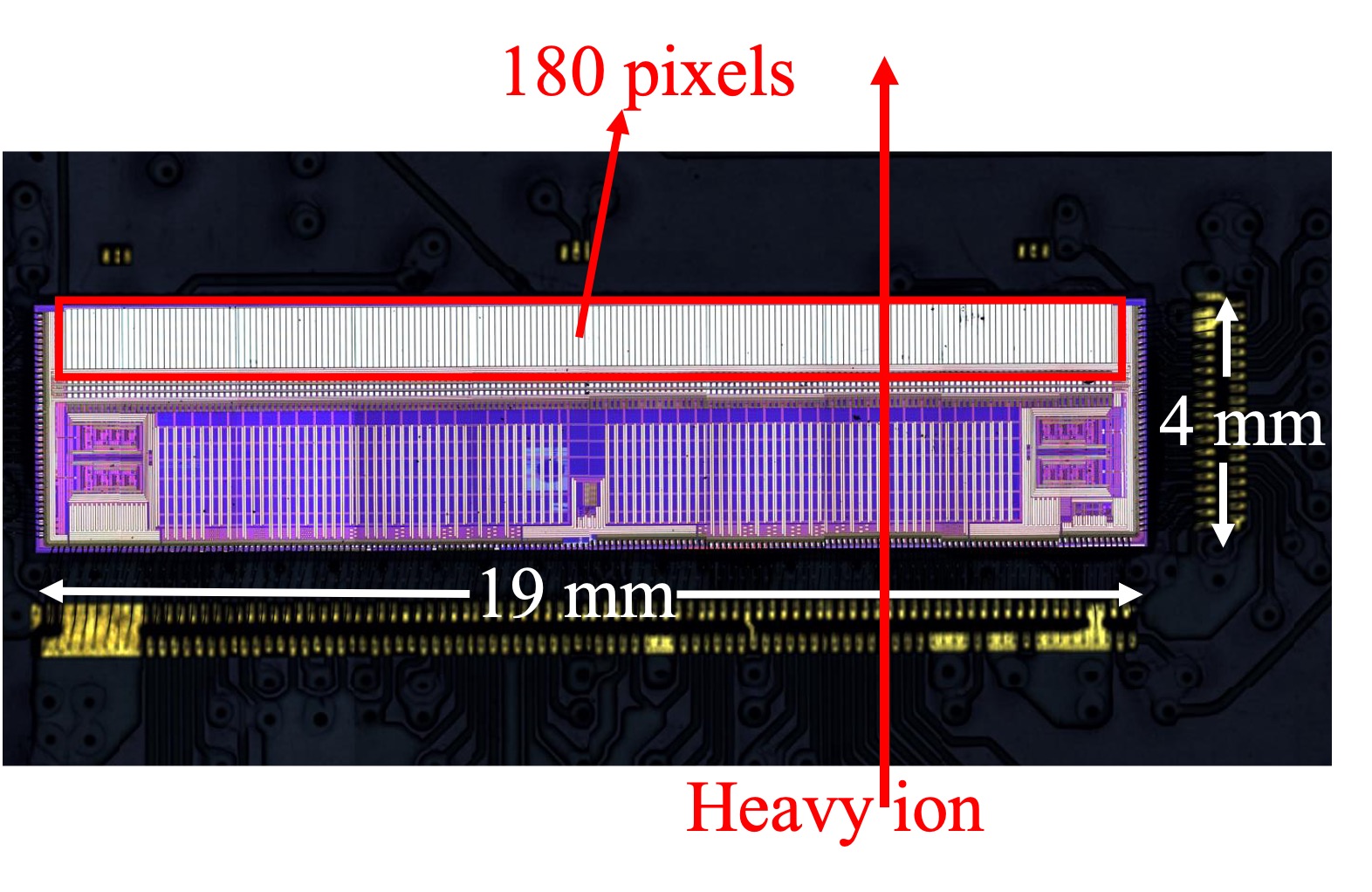}
    \quad
    \includegraphics[width=0.15\linewidth]{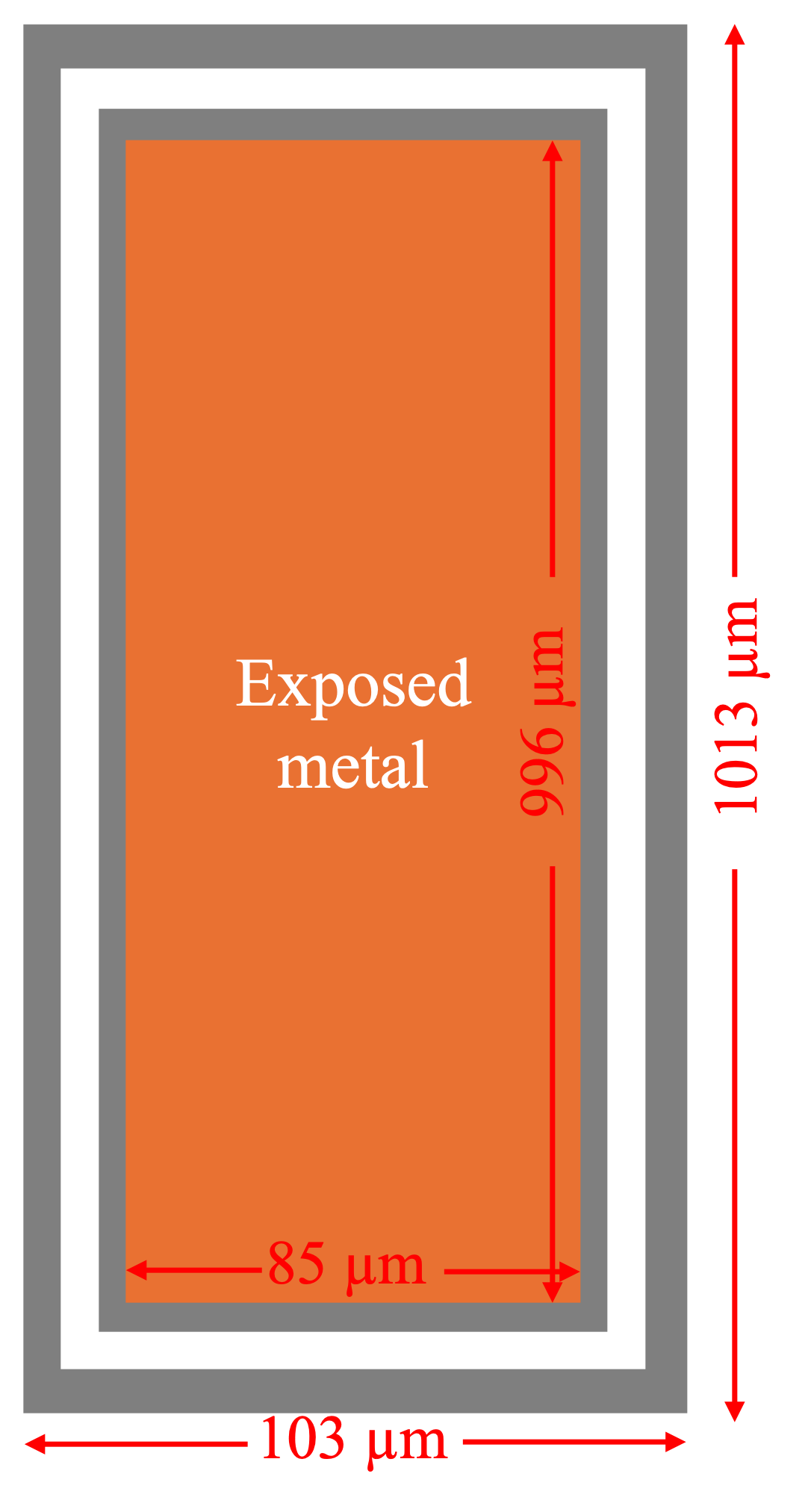}
    \caption{\label{fig:chip} A photograph of one Topmetal-CEE chip (left), and the schematic of the top of one pixel (right).}
\end{figure}

In each micro-TPC, four Topmetal-CEE chips are positioned in three rows in the bonding board, 
covering parts of the GEM area, as shown in Figure~\ref{fig:det} (right).
The data from the four chips are packaged in the front-end card, and then transmitted to the readout and control unit using a Samtec cable of about 10 m.
A photograph of one Topmetal-CEE chip is shown in Figure~\ref{fig:chip} (left).
Its size is 19 mm $\times$ 4 mm, containing a row of 180 pixels.
The lengths of the long and short edges of the pixel are 1 mm and 0.1 mm, respectively.
The detector is positioned so that the direction of the beam particle is parallel to the long edge.

The top of each pixel is an exposed metal surrounded by a guard ring, as schematically shown in Figure~\ref{fig:chip} (right).
The drifting electrons induce the current signals in the metal.
A charge-sensitive amplifier, discriminator, and logic circuit 
provide the amplitude measurement using the Time-Over-Threshold method, as well as the time measurement using an 8-bit counter with 25 ns precision.
The chip adopts a data-driven readout scheme with a speed of 40 MPixels/s.

A few drawbacks were identified in the first-generation chip~\cite{LIU2023167786}, 
and several improvements were implemented in the second-generation one and verified in the lab.
Both are fabricated in the GSMC 130 nm CMOS process.
The minimum operating threshold of the pixel is $\sim$ 20k $e^{-}$ for the first-generation chip due to the disturbance of the digital circuit to the analog circuit, 
while it is lowered to $\sim$ 5k $e^{-}$ for the second-generation one. 
The shaping time of the analog front-end has been improved from $\sim$ 1 $\mu$s to $\sim$ 0.5 $\mu$s.
For both generations of chips, the temporal noises are $\sim 350$ $e^{-}$, and input dynamic ranges of $> 100$k $e^{-}$ have been demonstrated.

\section{Performance of the prototypes}
\label{sec:perf}

The prototype using first-generation chips and single-stage GEM was assembled and tested with laser and heavy-ion beams,
to assess its performance of timing and position measurements, respectively.

\subsection{Laser beam test}
\label{subsec:laser}

The laser beam was used to simulate the track in the field cage with known drift distances.
The pulsed laser beam with a wavelength of 266 nm is generated by the Quantel Q-smart laser, and is focused and directed into the beam monitor via the laser optics,
with the direction orthogonal to the electric fields. 
The laser also provides a trigger signal which marks the start time of the event, and there is a delay below 1 $\mu$s between the trigger signal and the pulsed laser.

\begin{figure}[htbp]
    \centering
    \includegraphics[width=0.42\linewidth]{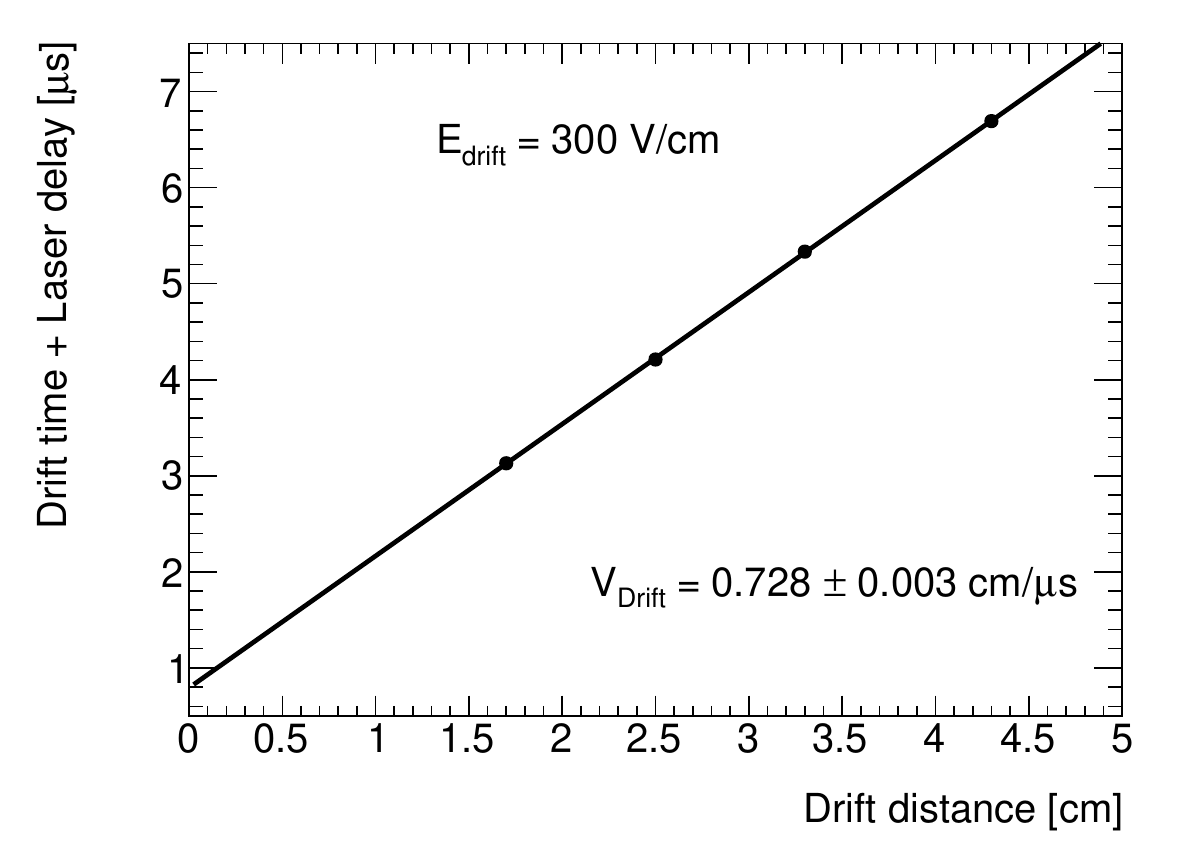}
    \quad
    \includegraphics[width=0.42\linewidth]{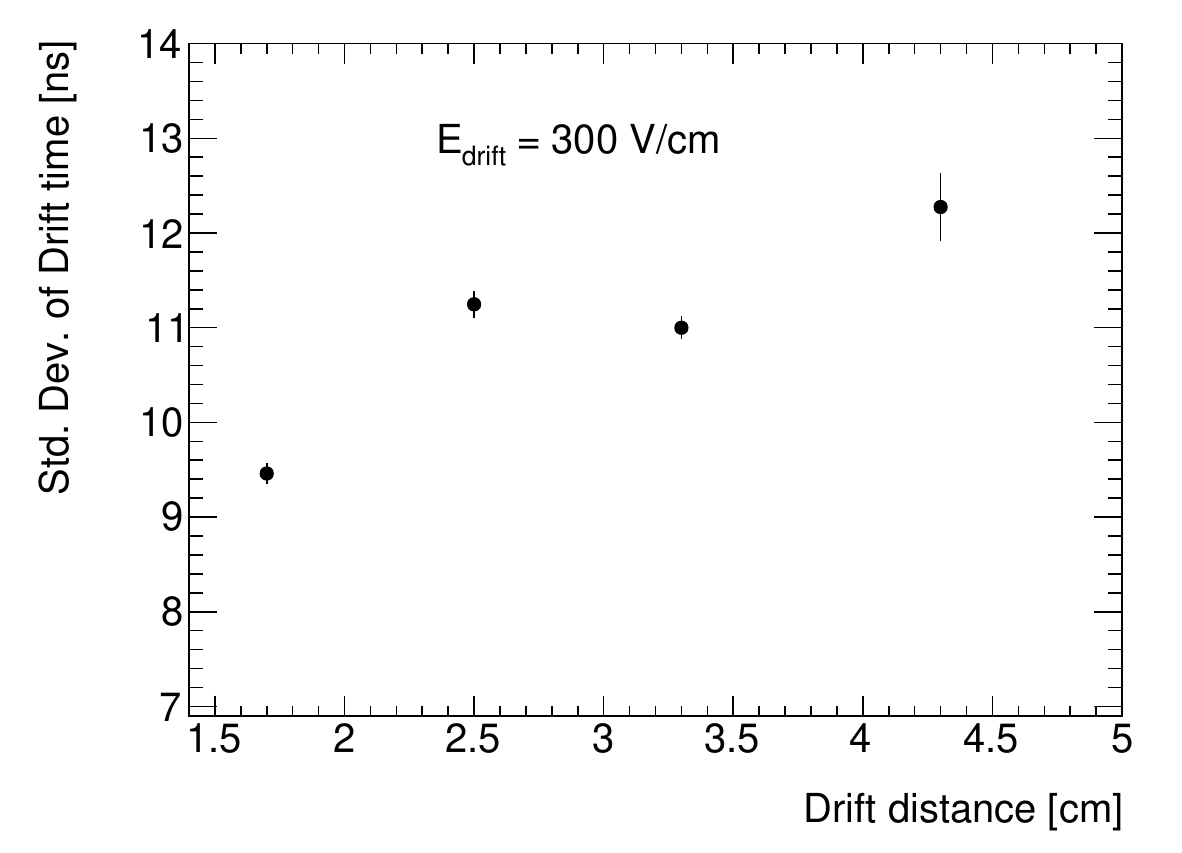}
    \caption{\label{fig:laserresult} The mean (left) and standard deviation (right) of drift time as a function of drift distance. }
\end{figure}

The test was conducted using the gas mixture of Ar(70\%) + CO$_{2}$(30\%) at room temperature and about local atmospheric pressure in Lanzhou.
The drift electric field was 300 V/cm. The $\mathrm{V_{GEM}}$ was 350 V, and the induction electric field was 1000 V/cm.
The pixel threshold was set to about 22k $e^{-}$. 
The average drift time as a function of drift distance is shown in Figure~\ref{fig:laserresult} (left).
A linear fit is used to retrieve the electron drift speed of 0.728 $\pm$ 0.003 cm/$\mu$s, which is consistent with the prediction.
The standard deviation of the drift times (shown in Figure~\ref{fig:laserresult} (right)), which reflects the time resolution of the beam monitor, increases from about 9 ns with the drift distance of 1.7 cm to about 13 ns with the drift distance of 4.3 cm.


\subsection{Heavy-ion beam test}
\label{subsec:ion}

The response of the beam monitor to the heavy-ion beam was investigated at HIFRL-CSR, on the site for the eventual CEE spectrometer.
The Kr-ion beam with an energy of about 320 MeV/u was used.
The detector configuration was the same as described in Section~\ref{subsec:laser}.

\begin{figure}[htbp]
    \centering
    \includegraphics[width=0.48\linewidth]{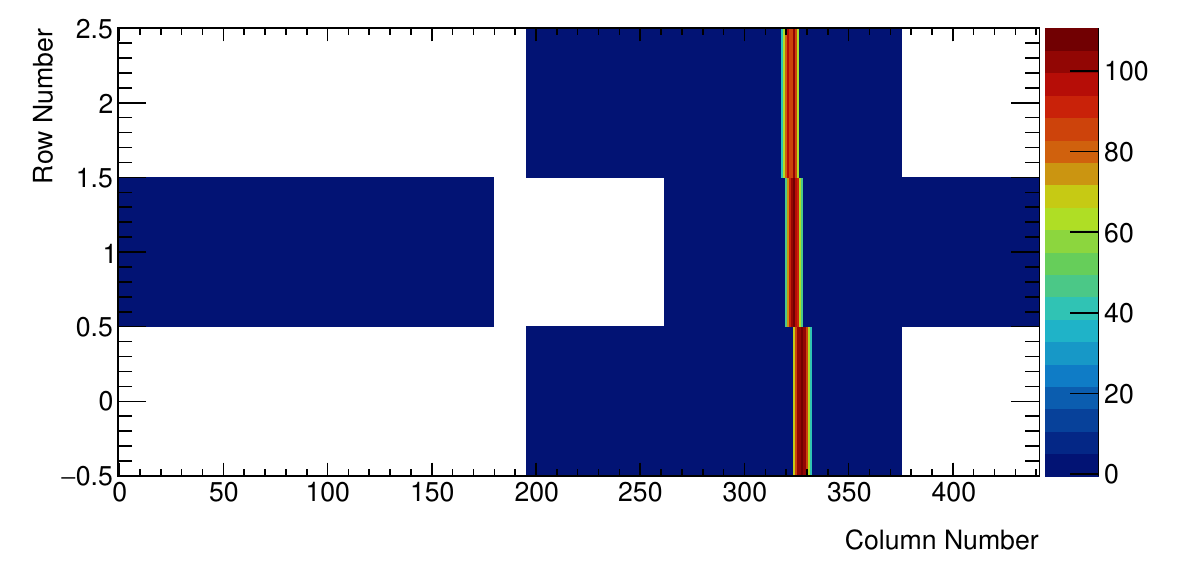}
    \quad
    \includegraphics[width=0.48\linewidth]{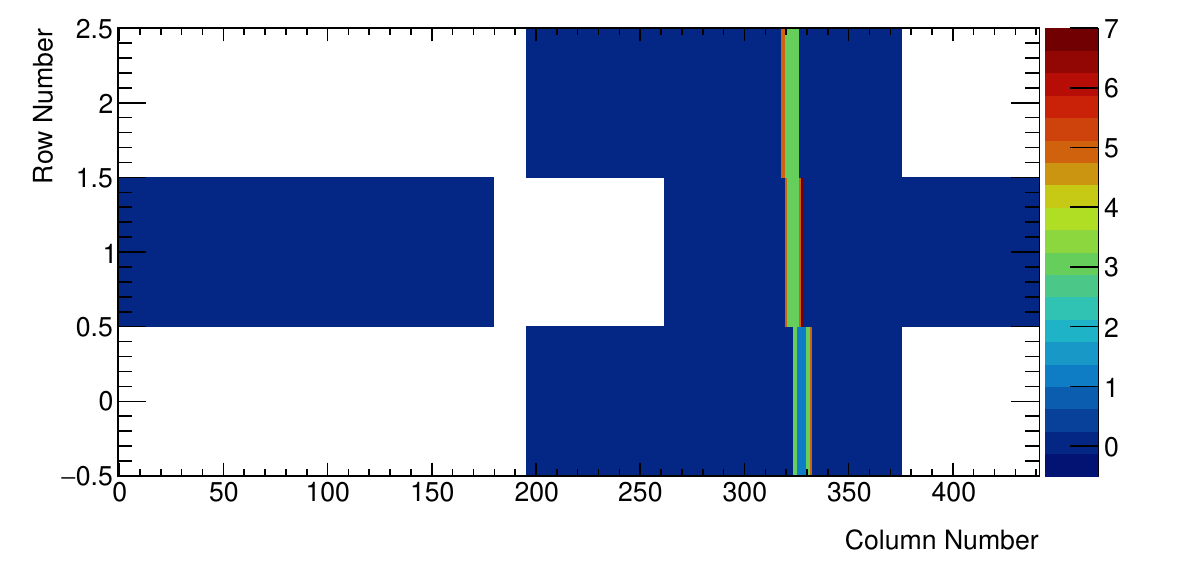}
    \caption{\label{fig:iontrack} A Kr-ion track with the color representing the amplitude (left) and arrival time (right). 
             In each plot, three rows of chips are represented by the row number, while the pixels in each chip are represented by the column number.}
\end{figure}

Figure~\ref{fig:iontrack} shows a typical Kr track observed in one of the micro-TPC of the beam monitor,
with the color representing the amplitude (left) and arrival time (right).
The unit of signal amplitude is about 700 $e^{-}$, while the unit of time is 25 ns.
For illustration purpose, the earliest arrival time is set to be 1 unit.
The white areas represent the places not covered by the pixels.
As expected, the pixels in the center of the clusters generally have larger signal amplitudes and earlier timestamps.

\begin{figure}[htbp]
    \centering
    \includegraphics[width=0.45\linewidth]{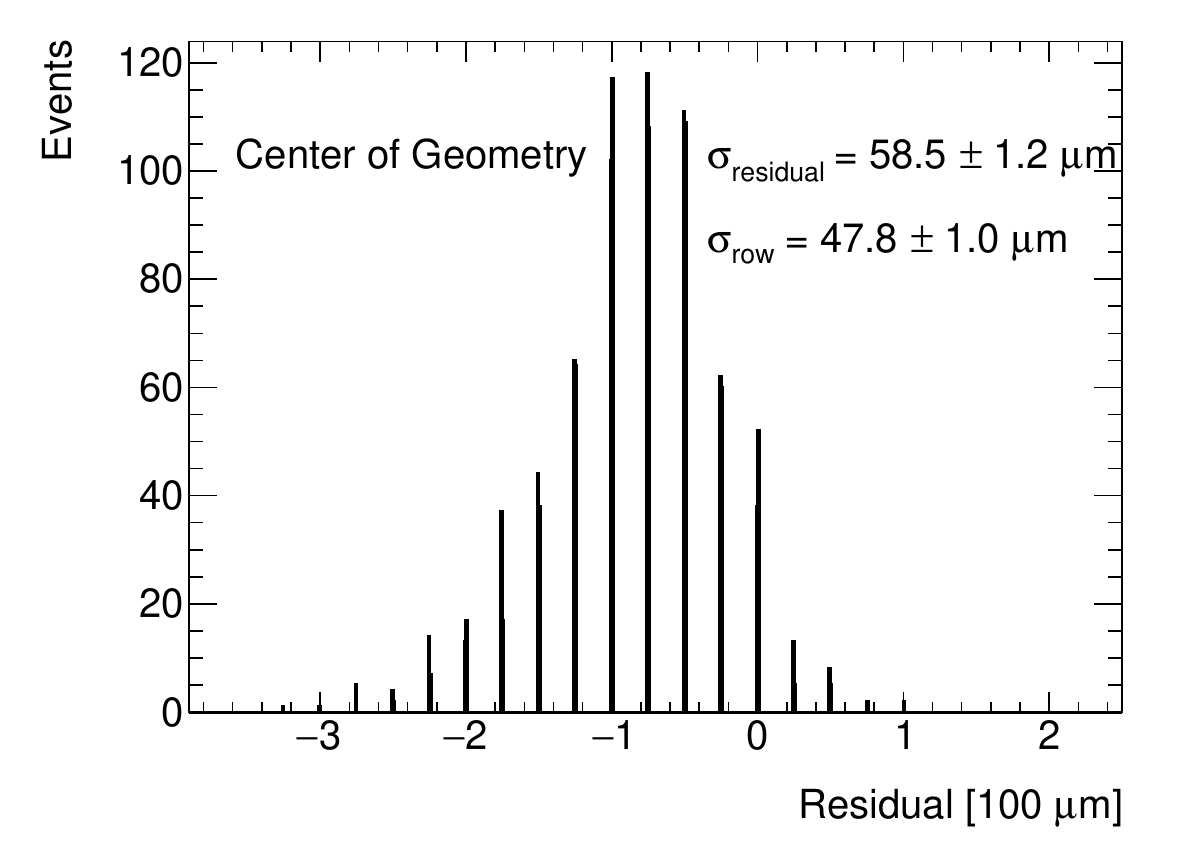}
    \quad
    \includegraphics[width=0.45\linewidth]{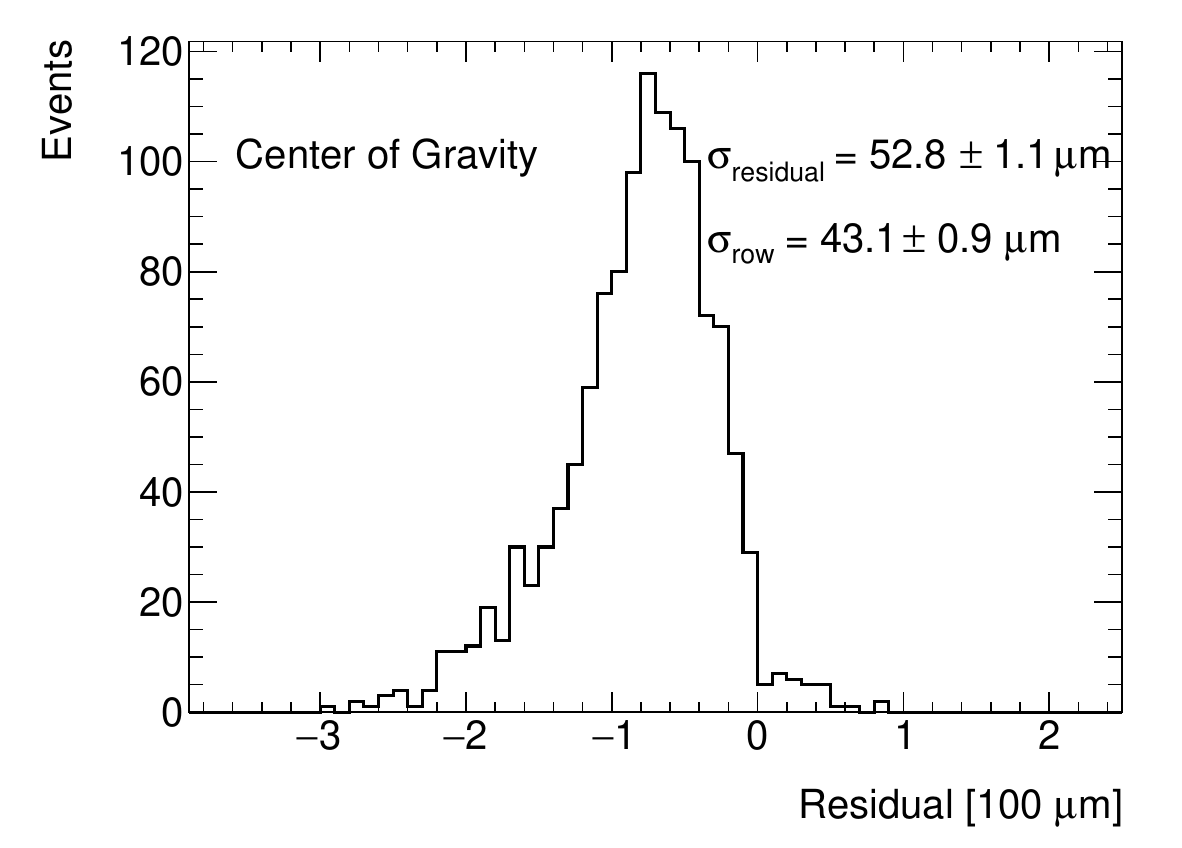}
    \caption{\label{fig:ionresidual} The residual distributions with the cluster positions determined using the center of geometry (left) and center of gravity (right) methods.
            As explained in the text, $\mathrm{\sigma_{residual}}$ is the standard deviation of the residual distribution, and $\mathrm{\sigma_{row}}$ is equal to $\mathrm{\sqrt{\frac{2}{3}}\sigma_{residual}}$. }
\end{figure}

The spatial resolution of single row of pixels, $\mathrm{\sigma_{row}}$, can be calculated using the tracks with clusters in all three rows.
The cluster positions in the 0th and 2nd row are used to define the track position, and the distance between the cluster position in the 1st row and the track defines the residual.
Assuming the spatial resolution of three rows of pixels are equal, $\mathrm{\sigma_{row}}$ is equal to $\mathrm{\sqrt{\frac{2}{3}}\sigma_{residual}}$,
where $\mathrm{\sigma_{residual}}$ is the standard deviation of the residual distribution.
Figure~\ref{fig:ionresidual} shows the residual distributions of 1241 tracks, with the cluster positions determined using the center of geometry method (left), i.e. geometrical average of pixel positions, and center of gravity method (right), i.e. signal weighted average of pixel positions.
The corresponding spatial resolutions are 47.8 $\pm$ 1.0 $\mu$m and 43.1 $\pm$ 0.9 $\mu$m, respectively.

\section{Conclusions}
\label{sec:con}

The beam monitor for the CEE experiment at HIFRL-CSR is under development.
Its main purpose is to measure the position and time of each beam particle, improving the precision of primary vertex reconstruction.
Custom-designed Topmetal-CEE pixel sensors are used as the direct anode readout, coupled with GEM for gas amplification.
Two generations of the chips have been produced, with the second generation having improved minimum operating threshold and shaping time.
The prototype using first-generation chips and single-stage GEM has been assembled and characterized.
The time resolution, which was obtained from the laser beam test, ranges from 9 to 13 ns as the drift length ranges from 1.7 to 4.3 cm.
The spatial resolution of 43 $\mu$m for one row of pixels was obtained from the heavy-ion beam test.
Prototypes with second-generation chips and with double-stage GEMs are being developed, and will be characterized using similar methods.

\acknowledgments
This work was supported by the National Natural Science Foundation of China (No. 11927901, 12105110, U2032209, 12275105),
the National Key Research and Development Program of China (No. 2024YFA1610700),
and Guizhou Science and Technology Foundation-ZK [2023] General (No. 248).


\bibliographystyle{JHEP}
\bibliography{CeeBMPixel2024Proceeding.bib}

\end{document}